\documentclass[preprint,aps,prb,endfloats]{revtex4}

\usepackage{epsfig}

\begin{document}

\title{Density functional study of alkali metal atoms and monolayers on
graphite (0001)}

\author{K. Rytk\"onen, J. Akola, and M. Manninen}
\affiliation{Nanoscience Center, Department of Physics, P.O. Box 35,
FI-40014 University of Jyv\"askyl\"a, Finland}

\date{\today}

\begin{abstract}

Alkali metal atoms (Li, Na, K, Rb, Cs), dimers and (2$\times$2)
monolayers on a graphite (0001) surface have been studied using
density functional theory, pseudopotentials, and  a periodic
substrate. The adatoms bind at the hollow site (graphite hexagon),
with Li lying closest to (1.84 \AA) and Cs farthest (3.75 {\AA})
from the surface. The adsorption energies range between
$0.55-1.21$~eV, and the energy ordering of the alkali adatoms is
Li$>$Cs$\ge$Rb$\ge$K$>$Na. The small diffusion barriers (0.02-0.21
eV for the C-C bridge) decrease as the atom size increases,
indicating a flat potential energy surface. The formation (cohesion)
energies of (2$\times$2) monolayers range between 0.55-0.81 eV,
where K has the largest value, and increased coverage weakens the
adsorbate-substrate interaction (decoupling) while a two-dimensional
metallic film is formed. Analysis of the charge density
redistribution upon adsorption shows that the alkali metal adatoms
donate a charge of $0.4-0.5 e$ to graphite, and the corresponding
values for (2$\times$2) monolayers are $\sim 0.1 e$ per atom. The
transferred charge resides mostly in the $\pi$-bands (atomic 
$p_z$-orbitals) of the outermost graphene layer.

\end{abstract}

\maketitle


\section{Introduction}

The adsorption of alkali metals on a graphite surface (highly
oriented pyrolytic graphite, HOPG) is widely studied for several
reasons. Firstly, adsorbed alkali metal adatoms (dispersed phase)
have shown a substantial activity in gasification reactions
(catalysis), \cite{Hoc93,Jan93,Lam98} and improved hydrogen
physisorption on graphitic hosts (HOPG, carbon nanotubes) has been
suggested.\cite{Cab05} Secondly, alkali metal monolayers (MLs)
exhibit interesting metallic properties as they appear as nearly
ideal two-dimensional (2D) quantum wells, in which the metal valence
electrons are confined and form discrete quantum well
states.\cite{Bre01,Bre02,Bre03,Alg06} Thirdly, most alkali metal
adatoms intercalate readily between graphene (GR) layers, and
technological applications of lithium-graphite intercalation
compounds have been introduced as rechargeable solid-state Li-ion
batteries. \cite{Dre81,Hu84,Jun89,Tar01}

Alkali metals show intriguing structural phase transitions on HOPG,
and - despite the similarities in their electronic structure - they
exhibit different properties as the adsorbate coverage is increased.
A common feature is the formation at low densities of a dispersed
and highly polarized phase with a maximal adatom-adatom distance
(``correlated liquid''). As the adatom coverage is increased a
critical density is obtained, after which a nucleation to more
closely packed configurations (islands) occurs.\cite{Hun98,Car05}
The differences between alkali metals arise in the island formation:
Li, for example, forms incommensurate hexagonally close-packed
islands on HOPG (provided that it does not intercalate via defects),
while K follows a (2$\times$2) construction. The larger alkali
metals (Rb and Cs) have been observed to form (2$\times$2)
overlayers, and ($\sqrt3\times\sqrt3$)R30$^\circ$, rotated
incommensurate hexagonal (2$\times2$)$^*$ as well as
($\sqrt7\times\sqrt7$)R19$^\circ$ phases have been reported for Cs.
\cite{Hu86,Hun96,Car05} Sodium is unusual, because its growth scheme
changes at 110~K from layer-by-layer to three-dimensional (3D)
clustering, and 3ML thick bcc(110) microcrystals with slightly
buckled surfaces have been observed.\cite{Bre01,Bre03} Unlike the
other alkali metals, Na also does not form stage 1 intercalation
compounds. Only stage 8 compounds have been
reported.\cite{Joh86,Car05}

The fundamental question of the alkali-HOPG systems concerns the
nature of the adsorbate-substrate interaction. It is widely held
that the alkali metal donates charge to the substrate $\pi$-bands,
which dominate the electronic band structure near the Fermi energy
(semimetal), \cite{Cha91,Cha92,Boe97} but the amount of charge
transferred is still unclear. Different experimental techniques
(electron spectroscopy, work function measurements, photoemission
and photoabsorption) as well as different theoretical approaches
(band structure calculations, cluster models) have led to
controversial results for the K-HOPG systems. As summarized in the
review article by Caragiu and Finberg, \cite{Car05} the charge
transfer in the dispersed phase spans a range of $0.3-0.7 e$ per
adatom, while the decoupled (2$\times$2) islands exhibit values
between $0.17-0.46 e$ per adatom. Experimental studies have focused
mainly on the adsorption of potassium, with much less data on the
charge transfer for other elements. Theoretical investigations of
these systems have focused on Li, Na, and K adatoms,
\cite{Lam98,Anc93,Whi94,Hjo98,Lou00,Gal03,Kha04,Zhu04,Ryt04,Piv05,Cab05,Val06}
while results for the (2$\times$2) coverage have been reported for
Li and K. \cite{Lam98,Anc93,Whi94,Hjo98,Kha04}

We report here a systematic density functional study of alkali metal
(Li, Na, K, Rb, Cs) adatoms, dimers, and (2$\times$2) monolayers on 
HOPG. The calculations have been performed using a periodic
``slab model'' that mimics the real HOPG surface and its electronic
band structure, and we use an extensive basis set to describe the
subtle adsorbate-substrate interaction accurately. Although other
theoretical studies on Li, Na, and K adatoms have been
published,\cite{Zhu04,Val06} our work fills a gap in the theoretical
description of alkali metal-HOPG systems. This is a natural extension 
to our previous study of Na adatoms and clusters on HOPG,
where the experimentally observed clustering behavior of Na adatoms
was reproduced.\cite{Ryt04} By applying the same simulation approach
to the all alkali metal atoms, we can point out differences between
them and seek possible explanations. We report the optimized
geometries, energetics, and charge transfer of the alkali-HOPG
systems, and provide visualizations of the charge density
redistribution (difference) and electron localization function in
order to shed light on the adsorbate-substrate interaction. We also
discuss the peculiar adsorption properties of Na that appear to be
related to its atomic radius and ionization potential.

\section{Simulation methods}

The calculations were performed using the Car-Parrinello molecular
dynamics program, \cite{CPMD} which is based on density functional
theory (DFT). The electron-ion interaction is described by
nonlocal, norm-conserving, and separable pseudopotentials of the
form suggested by Troullier and Martins. \cite{TM91} For Li and Na,
only the 2s/3s valence electron is treated explicitly, while for K,
Rb, and Cs we include the semi-core (p-shell) electrons. The program
uses periodic boundary conditions and a plane wave basis with a
kinetic energy cutoff of 70 Ry. The generalized gradient-corrected
approximation of Perdew, Burke and Ernzerhof (PBE) is adopted for
the exchange-correlation energy $E_{xc}$. \cite{PBE96} The atomic
positions are optimized using a quasi-Newton approach (BFGS method)
\cite{BFGS} until all Cartesian components of the nuclear gradients
are below 1{$\times$}{$10^{-4}$} atomic units. The electronic
Hamiltonian is rediagonalized after each optimization step using the
Lanczos method, and a finite temperature functional (T = 1000 K) by
Alavi {\it et al.} \cite{Ala94} is used for the Kohn-Sham (KS)
single-particle state occupancies. This reflects the small energy
gap (band gap) between the occupied and unoccupied states of
graphite.

The substrate is modeled as a periodic slab of three graphene layers
with a stacking {\it ABA} (Bernal graphite), and it comprises 96 
fixed C atoms (32 in each layer). Our previous experience has shown 
that three GR layers are needed in order to obtain converged
results for the alkali atom adsorption if a substantial charge
transfer to the substrate occurs.\cite{Ryt04} The nearest-neighbor 
C-C distance is 1.421~{\AA}, and the interplanar distance is fixed 
to the experimental value of 3.354 {\AA}, since the PBE functional 
used does not describe weak dispersion forces well.\cite{Kga03} The 
model is similar to that used in our earlier work,\cite{Ryt04} except 
that the supercell symmetry is hexagonal, not orthorhombic. This is a 
natural choice for graphite, and it enables us to use fewer 
{\bf{k}}-points in the simulations. Earlier benchmark tests with an 
orthorhombic supercell indicated that a 2$\times$2$\times$1 
Monkhorst-Pack {\bf k}-point mesh \cite{Mon76} is needed to converge 
the atomic forces, and a 5$\times$5$\times$1 mesh is required for the 
energies.\cite{Ryt04} For the hexagonal supercell symmetry, a 
2$\times$2$\times$1 mesh is adequate for both forces and energy. The 
lateral dimension of the hexagonal supercell is 9.84~\AA, and the 
perpendicular box size varies between $16.7-20.7$~\AA, depending on 
the adsorbate (the distance between the replicated slabs is $10-14$~\AA). 
The maximum lateral separation of the repeated adsorbates is 9.84~\AA, 
and the configuration referred to as a separated adatom corresponds to 
a (4$\times$4) monolayers.

The effect of the substrate relaxation has been tested in the case
of Li adatom by releasing either (a) the six nearest C atoms when Li
is above the hollow site (in the middle of a graphite hexagon) or
(b) the four nearest C atoms when Li is above the $\alpha$-site
(above a C atom). In the former case, the adatom rises 0.01~\AA, and
the C atoms move by 0.04~\AA. The total energy of the relaxed system
is only 0.006~eV lower than for the fixed substrate. In the latter
case, Li is lowered by 0.04~\AA\ as the C atoms move 0.01-0.04~\AA,
and the total energy is reduced by 0.01~eV. The changes will be even
less in the other alkali metals, where the binding is weaker.

Our previous investigations of Na-HOPG systems \cite{Ryt04}
indicated that one must use an extensive plane wave basis set in
order to describe the subtle charge transfer and redistribution. We
showed that the reduction in the cutoff from 70~Ry to 50~Ry led to
a 33\% (0.17 eV) weaker binding for Na and 7.8\% (0.19 \AA) larger
distance from the surface. The choice of $E_{xc}$ functional is also
important, because both the local density (LDA) and
gradient-corrected (GGA) approximations do not describe dispersion
forces reliably. While LDA often yields overbinding for the
metal-adsorbate interaction, GGA methods display the lack of
dispersion forces at a large separation between the graphite layers.
A benchmark calculation for a Li$\cdot$C$_{\rm 6}$H$_{\rm 6}$
complex results in values 0.22 eV (binding energy) and 2.03 {\AA}
(Li separation from the center of the benzene ring), whereas the
corresponding second-order M\o{}ller-Plesset (MP2) results are 0.25
eV and 2.25 {\AA}.\cite{Vol02} A DFT study by Valencia {\it et al.}
reported values 0.71/0.21 eV and 1.65/1.90 {\AA} for L(S)DA/PBE
functionals which emphasize the overbinding character of LDA.
Furthermore, we have performed another benchmark calculation for
K$^+\cdot$C$_{\rm 6}$H$_{\rm 6}$ in order to test a system with a
cation/$\rm\pi$ interaction.\cite{Ma97} The results for the K$^+$
binding energy and distance are 1.04~eV and 2.96~\AA, respectively,
compared to the MP2 results 0.74~eV and 2.81~\AA. \cite{Tsu01} The
experimental binding energy (0.83~eV) \cite{Sun81} suggests that PBE
slightly overbinds systems with pure cation/$\rm\pi$ interaction. We
note, however, that the interaction in alkali-benzene systems
differs from that on HOPG.

\section{Results}

The results for the alkali metal adatoms, dimers, and (2$\times$2)
monolayers on HOPG are presented in Table \ref{tab1}. The formation
(cohesion) energy ($\Delta E$), defined as the energy difference
between the system and its constituents (separated metal atoms and
substrate), comprises the adsorption energy ($\Delta E_\bot$) and
the binding energy ($E_b$) of a separated adsorbate (per atom). The
potential energy surface of a single alkali adatom (Li, Na, K, Rb,
Cs) has been mapped by optimizing the distance from the graphite
(0001) surface at four locations: above the hollow site (hexagon
center), above the $\alpha$ and $\beta$ sites (carbons), and above
the bridge site (C-C bond). For each alkali adatom, the hollow site
is favored energetically, and the optimal diffusion path from one
minimum to another one is via the bridge site. The corresponding
diffusion barrier height $E_{diff}$ is significantly larger for Li
than for the other alkali metals ($0.02-0.06$~eV, see Table \ref{tab1}).
\cite{PES} These results agree with the findings of Lamoen and Persson, 
who obtained a maximum variation of 0.05 eV for the K adsorption at
different locations. \cite{Lam98} The preference of alkali metals
for the hollow site is well known.
\cite{Anc93,Lou00,Kha04,Zhu04,Zhu05}

\begin{figure}
\epsfig{file=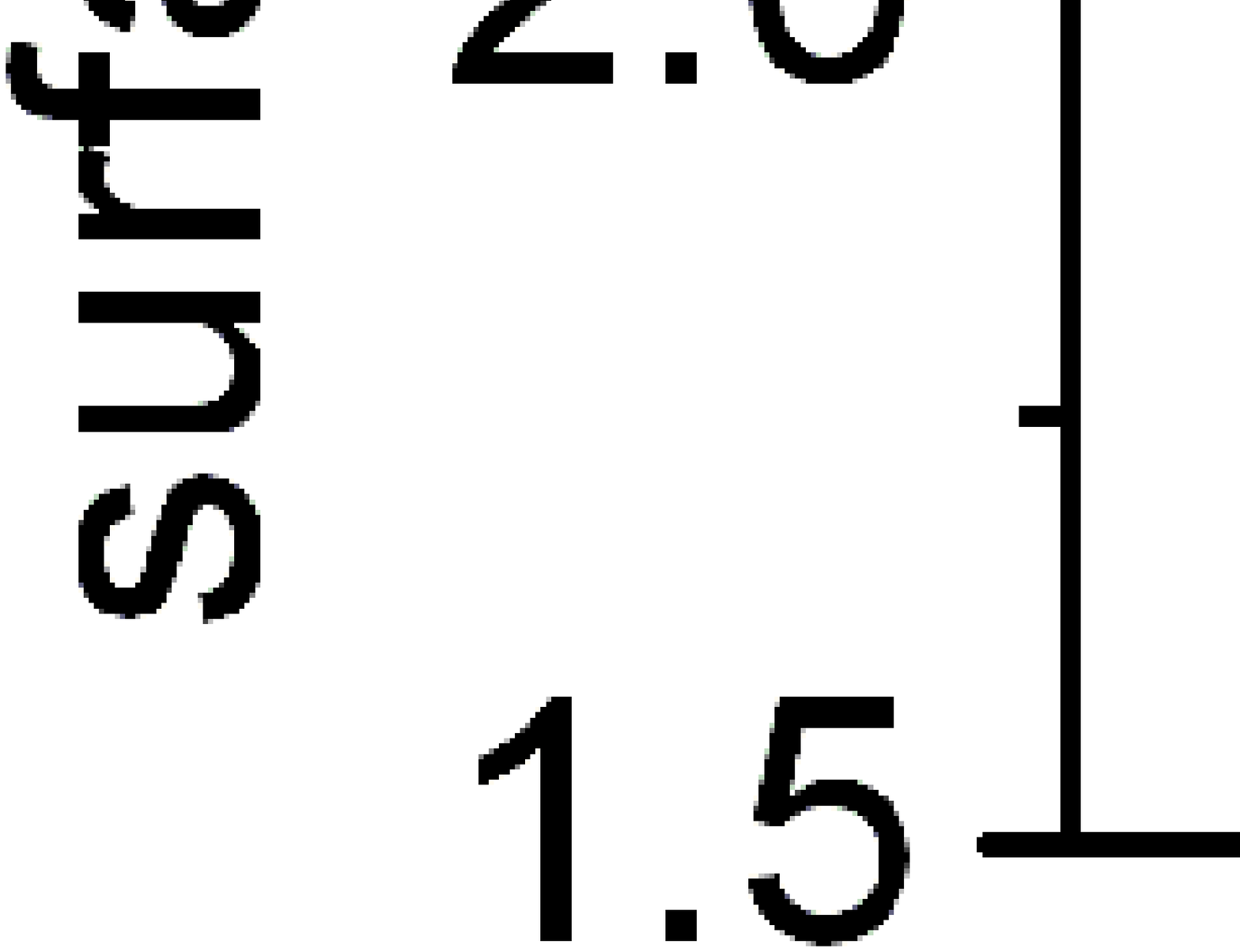, scale=0.11}
\caption{Adsorption of alkali metal atoms and (2$\times$2) monolayers on HOPG:
(a) the formation and adsorption energies per atom ($\Delta E$ and $\Delta E_\bot$),
and (b) the vertical separation from the substrate (d$_\bot$).}
\label{trend}
\end{figure}

Fig. \ref{trend} shows the formation and adsorption energies
($\Delta E$ and $\Delta E_{\bot}$) as well as the separation from
the surface ($d_\bot$) for alkali metal adatoms and (2$\times$2)
monolayers (see also Table \ref{tab1}). In the case of adatoms, the
adsorbate binding energy is negligible, and the adsorption energy is
equal to the formation energy. The $\Delta E$ values of Li and Na
differ, since Li binds relatively strongly to the substrate (1.21
eV), whereas Na has the weakest adsorption (0.55 eV) of all the
adatoms. The three larger alkali atoms (K, Rb, Cs) have $\Delta
E$/$\Delta E_\bot$ values of the same magnitude (0.99-1.04 eV). For
the (2$\times$2) monolayers, the trend in $\Delta E_\bot$ is similar
to that in separated adatoms, with Li and Na providing the
upper/lower boundaries (0.39/0.16~eV). The behavior of the formation
energies is slightly different, as K forms the most stable
(2$\times$2) construction, and $\Delta E$ decreases for Rb and Cs.
Despite the fact that Li and Na do not form (2$\times$2) monolayers,
Li also has a considerable formation energy. Cesium becomes slightly
compressed in the 2D graphite mesh (nearest-neighbor distance 4.92
{\AA} compared to 5.24 {\AA} of bulk Cs), causing a lower adsorbate
binding energy. The adatoms (especially K) have larger $\Delta E$
values than (2$\times$2) monolayers, which supports the experimental
observation of the dispersed phase at low adatom coverage.

The surface separation of alkali metal adatoms [Fig. \ref{trend}(b)]
increases as the atomic radius increases, and the $d_\bot$ values
range between $1.84-3.44$~\AA. For the (2$\times$2) monolayers,
$d_\bot$ increases systematically as the metal films decouple from
the surface, and the corresponding range of values is
$2.02-3.75$~\AA. The decoupling effect is reflected in the
adsorption energies, which are smaller than for the separated
adatoms. The Na (2$\times$2) monolayer is remarkably far from the
surface (3.16~\AA), illustrating further that this construction is
unfavorable. Among the two smallest alkalis, Li appears more able to
adjust to the underlying graphite lattice. This is in accordance
with the experimentally observed intercalation properties.

An exceptionally weak interaction of Na$_2$ with HOPG was found in
our earlier study, \cite{Ryt04} and further calculations of both
horizontally and vertically aligned dimers show that the adsorption
is also weak for the latter orientation (Table \ref{tab1}). This
indicates that Na$_2$ is almost decoupled from the surface, and the
corresponding formation energy ($\Delta E$) is similar to that of a
single adatom. Alkali metal dimers have two valence electrons and a
closed-shell electronic structure, and their interaction with HOPG
is significantly weaker than for adatoms. Comparison of the dimer
bond lengths ($R_{dim}$) shows that Li$_2$ and K$_2$ elongate
(10-15\%) upon binding with HOPG, with a smaller effect in Na$_2$.

In the case of two K (2$\times$2) overlayers, $\Delta E_\bot$ is
approximately one half of the corresponding value for one ML (Table
\ref{tab1}), but the total adsorption energies are close ($\sim$1.0
eV per substrate area). The surface separation of the lower
monolayer ($d_\bot$=3.11~\AA) is also similar to that for one ML
($d_\bot$=3.17~\AA), so that the interaction between the K
overlayers and HOPG is weak. The experimental value
$d_\bot$=2.79$\pm$0.03 {\AA} for a K (2$\times$2) layer (low-energy
electron diffraction, LEED) \cite{Fer04} is closer to the value
($d_\bot$=2.72~\AA) obtained for separated K atom (Table
\ref{tab1}). The separation of the K overlayers is 3.90 {\AA}, and
the nearest-neighbor distance between K atoms at different layers is
4.85 {\AA} (within a layer 4.92 \AA). The formation energies of 1ML
and 2ML systems are similar, and we conclude that K atoms bind with
0.8-1.0 eV on HOPG. The desorption kinetics model by Lou {\it et
al.} that is fitted to experimental data yields a desorption energy
of 1.0 eV.\cite{Lou00}

Our calculations of alkali metal adsorption are compared with other
DFT studies in Table \ref{tab2}. We have included results for Li,
Na, and K, but we found no theoretical studies of larger alkali metal atoms on
graphite. The range of adatom cohesion energies is $1.10-1.68$~eV
for Li, $0.50-0.72$~eV for Na, and $0.51-1.67$~eV for K, and our
values are near the lower bound for Li and Na. For K we obtain
$\Delta E$ that is in the middle of the broad range. The adatom
$d_\bot$ values show deviations of $0.2-0.4$~\AA, where our results
represent the upper boundary for Li and Na. There are two reasons
behind the scatter of data: LDA is known to overbind, and the
cluster models are not completely representative of HOPG because of
their electronic structure (energy gap between the occupied and
unoccupied orbitals). Furthermore, the good agreement with the
recent study by Valencia {\it et al.}\cite{Val06} is due to the
similar simulation model (slab geometry, PBE functional, plane wave
basis). We know of no other calculations of the Na (2$\times$2)
monolayer, and the comparison is limited to Li and K. For Li, our
results differ significantly from the study by Khantha {\it et al.},
\cite{Kha04} emphasizing the sensitivity to the choice of $E_{xc}$
functional. Benchmark calculations for a Li$\cdot$C$_{\rm 6}$H$_{\rm
6}$ complex have demonstrated that LDA leads to overbinding,
\cite{Val06} and we have more confidence in our PBE results. On the
other hand, for K (2$\times$2) ML the experimental layer spacing
(2.79 \AA, LEED)\cite{Fer04} is closer to the LDA result.

\begin{figure}
\epsfig{file=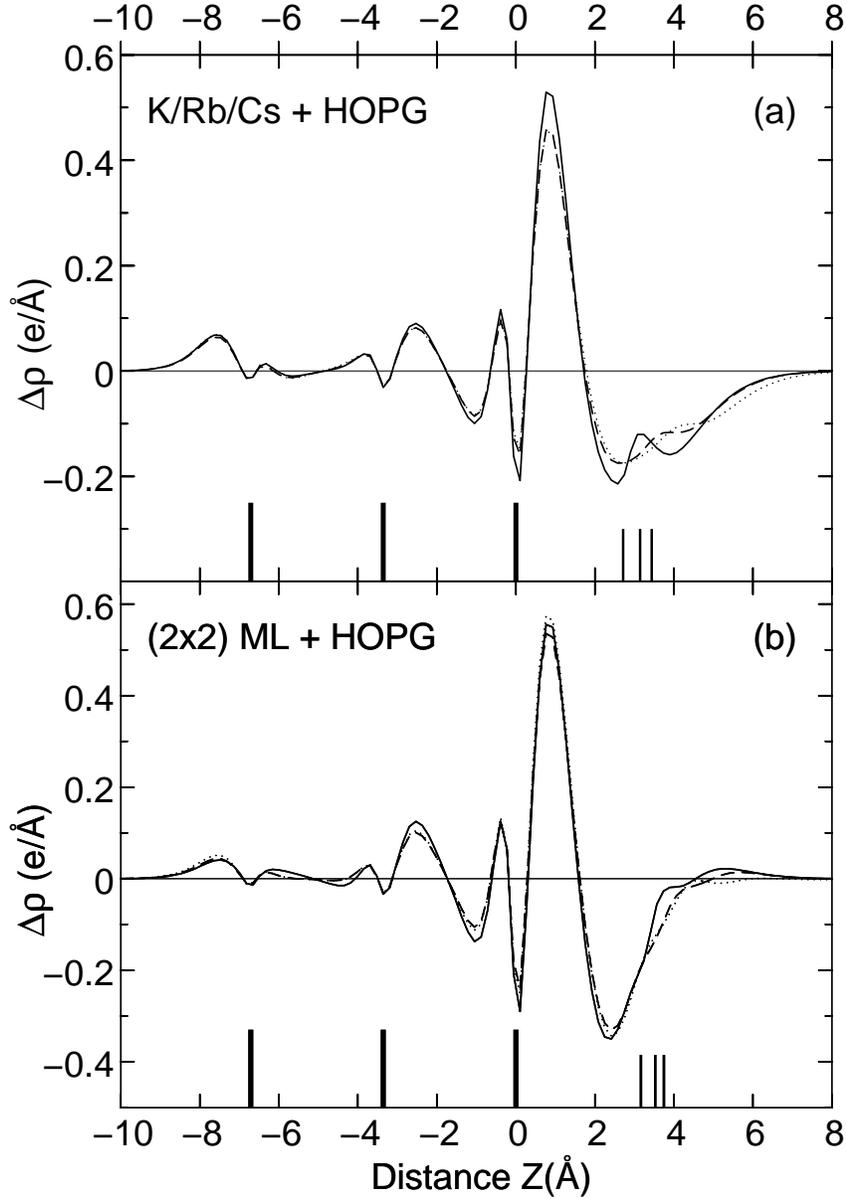, scale=0.65}
\caption{Laterally averaged charge density difference ($\Delta\rho_\bot$, {\it z}
direction) of K (solid curve), Rb (dashed curve), and Cs (dotted curve) adatoms
as well as (2$\times$2) monolayers on graphite. (a) Separated adatoms, and (b)
(2$\times$2) monolayers. The vertical bars denote the positions of GR layers
(thick bars) and alkali metal (thin bars). The charge densities have been
calculated in an extended simulation box, so that the distance between the
vertically repeated substrates is 20 {\AA}. Notice the different scale of
$\Delta\rho_\bot$ in (a) and (b).}
\label{K_Rb_Cs}
\end{figure}

Charge transfer between the adsorbate and substrate has been studied
by subtracting the calculated electron densities of HOPG and metal
layer from that of the whole system.\cite{Anc93,Lam98,Ryt04} 
Fig. \ref{K_Rb_Cs} shows the laterally averaged charge density 
difference ($\Delta\rho_\bot$, $z$ direction) for K, Rb and Cs adatoms 
and monolayers. In order to obtain reliable $\Delta\rho_\bot$ profiles,
we have recalculated the electron densities within a vertically
extended simulation box ($c$=26.71 {\AA}). A significant
accumulation below the lowermost graphene layer (GR1) was found for
Na in our earlier study ($c$=18.71 {\AA}), \cite{Ryt04} and we
confirm that this is a finite size effect caused by the
dipole-dipole interaction of the periodic systems.\cite{Val06} Our
benchmark calculation show that this undesirable effect of
periodicity is reduced as the perpendicular distance between the
slabs is increased up to 20 {\AA}.

Fig. \ref{K_Rb_Cs} shows that charge is depleted mainly from the
adsorbate and accumulated in the vicinity of the topmost graphene
layer (GR3). There is also a small accumulation on both sides of the
lower graphene layers (GR1 and GR2), and a small depletion within
the layers. There are only small qualitative differences between the
three largest alkali metals: the location and width of the charge
depletion node (alkali metal) depends on the atomic radius, and K
has the most pronounced accumulation above GR3. Within the substrate
range, the $\Delta\rho_\bot$ curves are identical. For the
(2$\times$2) MLs, the curves resemble closely each other, again,
with minor deviations near the adsorbate planes. The
$\Delta\rho_\bot$ variations are larger for MLs compared to adatoms,
and the weight of GR2 is slightly enhanced.

A layer-by-layer analysis of the charge transfer in the alkali-HOPG
systems is given in Table \ref{tab3}. Also shown is the
adsorbate-substrate cutoff distance (R$_{cut}$), which defines where
the charge accumulation chances to depletion above the substrate.
This distance is relatively insensitive to the atomic radius of the
alkali metal atom, and for adatoms its values range from 1.67~\AA\
(Li) to 1.79~\AA\ (Cs). The situation changes little for the
(2$\times$2) monolayers (Li and Cs 1.48 and 1.63 {\AA},
respectively), but there is a systematic shift downwards. The charge
transfer is calculated by integrating $\Delta\rho_\bot$ over a range
of $z$-values, and corresponds for a metal adsorbate to the negative
node (depletion area) around the alkali metal layer. With small
ionic radii and large ionization potentials (IP), Li and Na donate
small amounts of charge (${\Delta}q\sim -0.4 e$), while the other
adatoms show a depletion of $\sim 0.5$ electrons. The charge transfer
changes little when the coverage is increased from a
(4$\times$4) monolayer (separated adatom) to a (2$\times$2)
monolayer. The charge transfer from individual adsorbate atoms then
decreases as the alkali coverage increases, in accordance with
the calculated adsorption energies (Table \ref{tab1}). The largest
change is observed for Na, where ${\Delta}q$ changes from $-0.44$
to $-0.34 e$ ($-0.09 e$ per adatom) upon increasing the coverage.
This is further evidence of the unusually strong decoupling of Na.
Furthermore, inspection of the individual GR layers shows
that the largest charge accumulation occurs in the topmost layer
(GR3), and the net contribution of the two lower layers (GR1 and
GR2) ranges between 29-45\%.

\begin{figure}
\epsfig{file=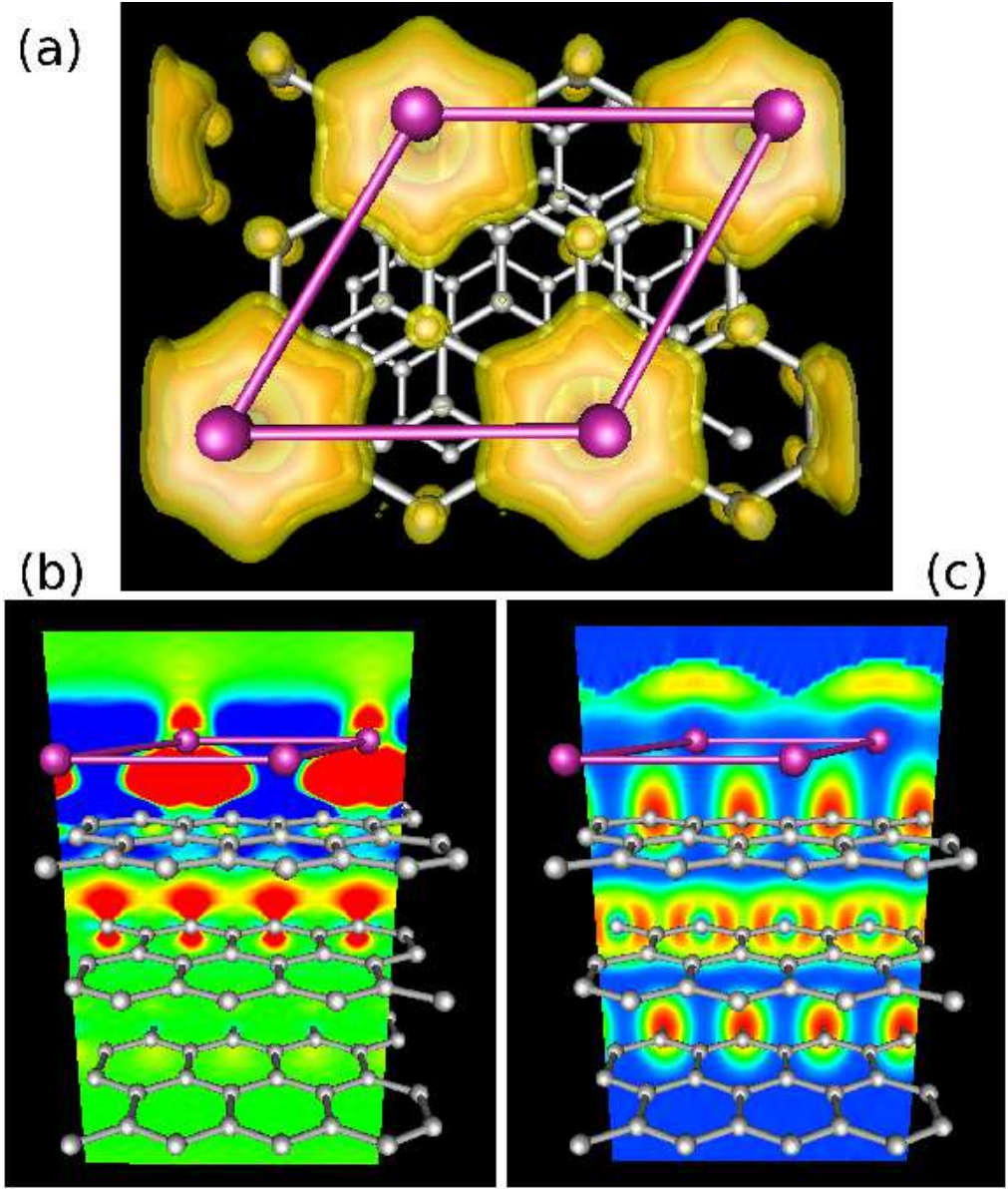, scale=0.23}
\caption{Visualization of the Li (2$\times$2) monolayer on HOPG. (a) Three
isosurfaces for the accumulated electron density. The corresponding values are
0.001 (yellow), 0.002 (orange), and 0.004$e$/{\AA}$^3$ (red), respectively.
The Li atoms are marked by magenta spheres. (b) Cutplane presentation of the
charge density difference ({\it xz} plane), where the red color corresponds to
accumulation (0.0005$e$/{\AA}$^3$ or more) and blue depletion (-0.0005$e$/{\AA}$^3$
or less). (c) The electron localization function (ELF, {\it xz} plane), where the
red color corresponds to full localization (1.0, covalent bonds), green is
analogous to homogeneous electron gas (0.5, metallic bonding), and blue equals
to low localization (0.0).}
\label{vis_LiML}
\end{figure}

The charge density difference ($\Delta\rho$) and electron
localization function (ELF)\cite{ELF} of Li (2$\times$2) ML on HOPG
are shown in Fig. \ref{vis_LiML}. This is not a stable
monolayer, as Li tends to form hexagonal incommensurate
structures (if it does not intercalate), but it serves to
visualize the Li-HOPG interaction. Each Li atom donates charge to
the six nearest C atoms [Fig. \ref{vis_LiML}(a)], and the underlying
hexagonal symmetry is reflected in the shape of the charge
accumulation lobes. Other areas of charge accumulation [Fig.
\ref{vis_LiML}(b), red color] can be seen above the Li atoms and
above and below the C atoms of GR2. Charge depletion (blue color) is 
visible in the Li layer, and especially between the Li atoms. The 
topmost graphene layer also shows significant depletion. The electron
localization function [ELF, Fig. \ref{vis_LiML}(c)] reflects the
probability of finding two electrons at the same location, and it
ranges from 0 (no localization, blue) to 1 (complete
localization, red). A reference value of 0.5 (green color)
corresponds to a homogeneous electron gas (metallic bonding), and
covalent bonds appear red due to the electron pairing (see C-C
bonds). The ELF shows no electron overlap within the Li ML, indicating 
weak chemical bonding between the Li atoms, and the very low 
electron density within the Li plane is consistent with 
the small value of ELF. This is reflected also in the
adsorbate-substrate interface, and the lack of electron overlap
suggest that there are no Li-C bonds present. However, the ELF
contours of the nearest C-C bonds deform towards Li, 
which can be regarded as a sign of polarization
(induction) that is an important component in the cation/$\rm\pi$
interaction.\cite{Tsu01} Finally, charge redistribution results in
a 'cap' of localized electron density above each Li atom.

\begin{figure}
\epsfig{file=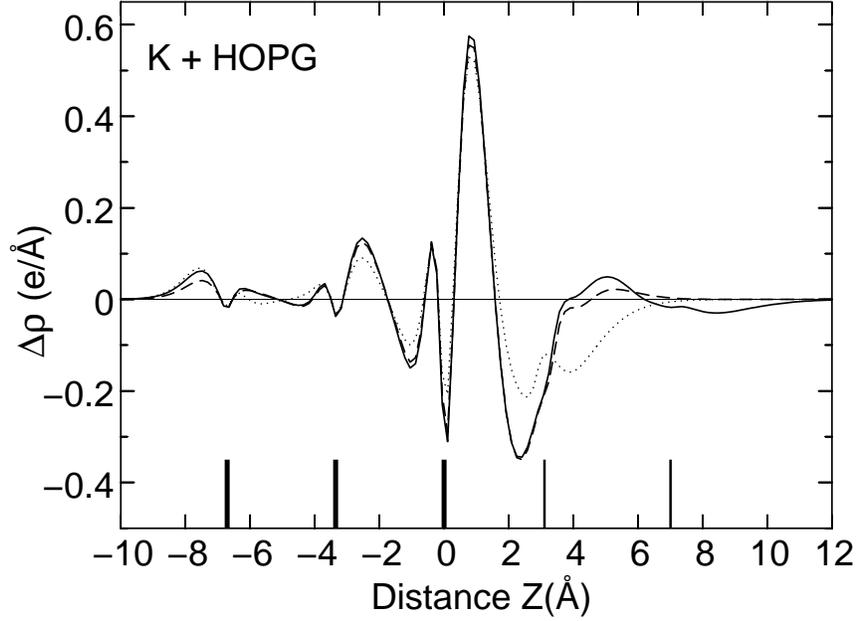, scale=0.65}
\caption{Laterally averaged charge density difference ($\Delta\rho_\bot$) of a K
adatom (solid curve), a (2$\times$2) monolayer (dashed curve), and two (2$\times$2)
overlayers (dotted curve) on graphite. The vertical bars denote the positions of
GR (thick bars) and K layers (thin bars, 2ML). The charge densities have been 
calculated in an extended simulation box, so that the distance between the vertically 
repeated substrates is 20 {\AA}.}
\label{K_curve}
\end{figure}

The laterally averaged charge density difference ($\Delta\rho_\bot$)
curves for a separated K adatom, a (2$\times$2) monolayer, and two
(2$\times$2) overlayers on HOPG are shown in Fig. \ref{K_curve}. 
Table \ref{tab3} shows that there is a charge transfer from the metal
film towards HOPG: ${\Delta}q = -0.53 e$ for a separated K atom, 
$-0.50 e$ for K (2$\times$2) ML, and $-0.48 e$ for two K (2$\times$2) 
layers. Decoupling causes visible differences: the charge is
depleted from the K adatom, whereas in the cases of (2$\times$2)
overlayers depletion is mainly beneath the lower K layer. The curves 
for 1ML and 2ML below the adsorbate are almost identical, the 
charge transfer per K atom in the adsorbate-substrate interface
is slightly smaller for 2ML ($-0.12 e$) than for 1ML ($-0.13 e$),
and there is a small accumulation between the K layers. Similar
analyses of $\Delta\rho_\bot$ in other DFT calculations have
given ${\Delta}q$ values ranging between $-0.38$ and $-0.46 e$ for
K, \cite{Lam98,Anc93,Val06} and $-0.17 e$ for (2$\times$2)
ML.\cite{Anc93,Lam98}

\begin{figure}
\epsfig{file=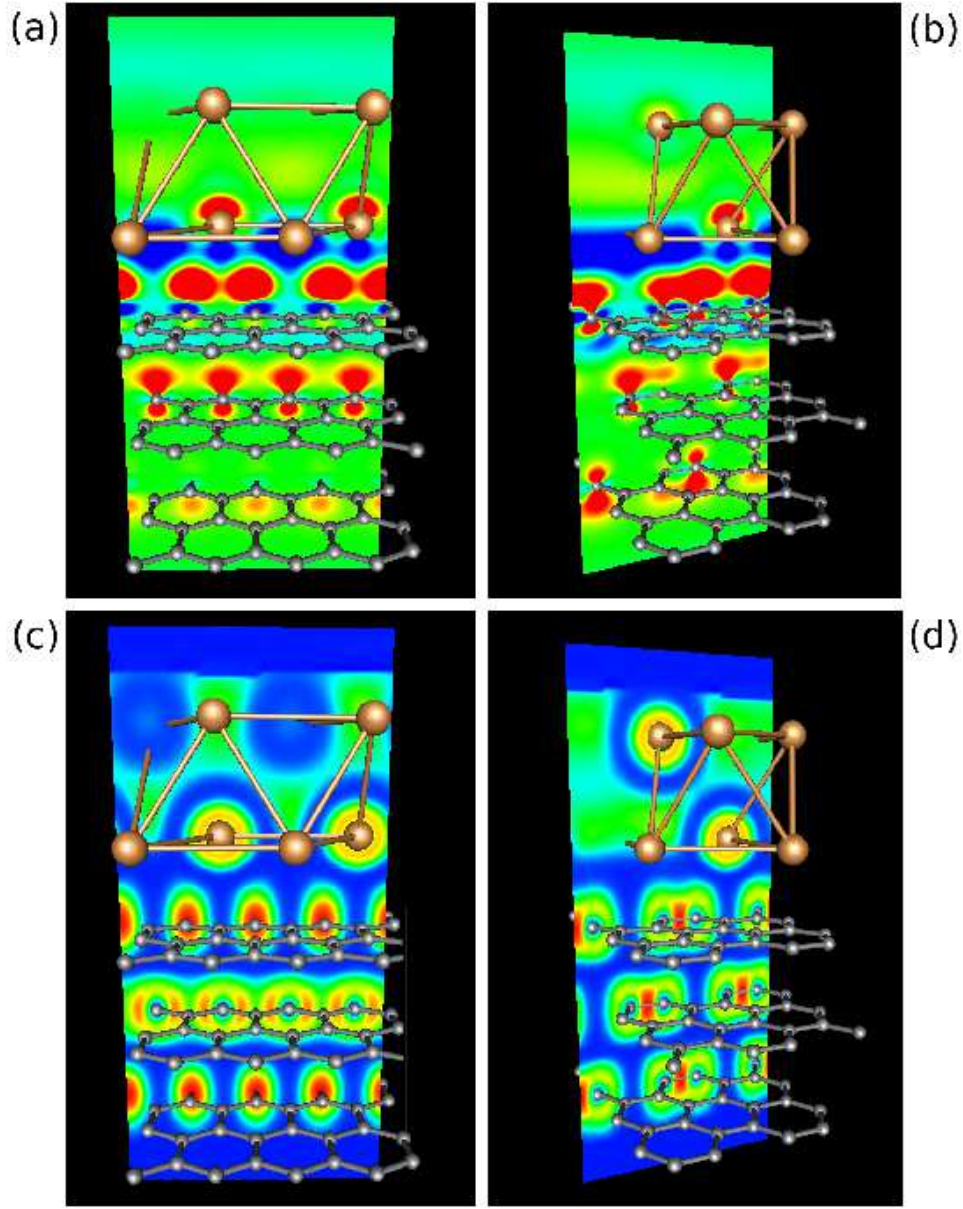, scale=0.23}
\caption{Cutplane visualization of two K (2$\times$2) overlayers on HOPG.
(a-b) The charge density difference is presented in {\it xz} and {\it yz} planes, 
where the red color corresponds to accumulation (0.0005$e$/{\AA}$^3$ or more) and 
blue depletion (-0.0005$e$/{\AA}$^3$ or less). (c-d) Similar presentation of ELF 
(see the caption in Fig. \ref{vis_LiML}).}
\label{vis_K}
\end{figure}

The charge density difference and ELF of two K (2$\times$2)
overlayers on HOPG are given in Fig. \ref{vis_K}. Charge 
accumulates mainly above the C atoms of the topmost GR layer [Figs.
\ref{vis_K}(a) and \ref{vis_K}(b)], and other regions can be seen above 
and below the C atoms of GR1 and GR2, and in the vicinity of K atoms.
Depletion occurs mainly below the lower K layer and in the C-C bonds
of GR3. Furthermore, a trace of depletion (light blue color) can be 
observed above the topmost K layer. These findings lead to a picture 
where the alkali metal donates charge to the $\pi$-bands (atomic
$p_z$-orbitals) of the GR planes, and there is depletion in the
$sp^2$ hybridized $\sigma$-bands of the topmost GR layer. The
electron localization function [Figs. \ref{vis_K}(c) and \ref{vis_K}(d)] 
shows that the electron density is delocalized ($\sim$0.5) between the K
atoms, so that K forms a metallic layer on HOPG. The boundary
between the metal film and 'vacuum' is sharp and flat, and this
is confirmed by the electron density contours (not shown). This
picture agrees with the He-scattering experiments for alkali
(2$\times$2) layers, where no corrugation was found in the surface
potential.\cite{Whi94} The blue rings around the individual K atoms
show that it is extremely unlikely to find more than one electron
within a range of the K $4s$ orbital (notice the localized $3p$ 
electrons). ELF shows that there is no chemical bonding between 
the adsorbate and substrate, as the interface region appears in 
blue color (delocalization). This indicates that the 
adsorbate-substrate interaction should be viewed as ionic.

\begin{figure}
\epsfig{file=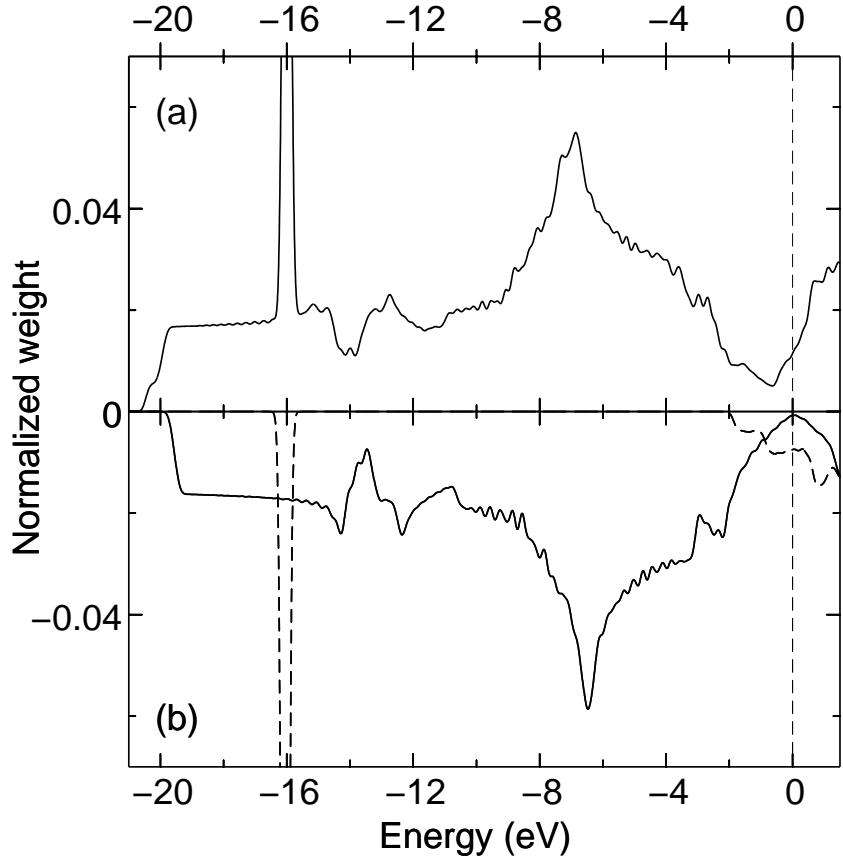, scale=0.65}
\caption{Normalized density of states (DOS) of (a) two K (2$\times$2) overlayers on
HOPG, and (b) the separated substrate (solid line) and K film (dashed line).
The energy bands are calculated with a 13$\times$13$\times$1 {\bf k}-point mesh, and
Gaussians of 0.10 eV width were used for each data point. The vertical dashed line
marks the Fermi energy.}
\label{DOS}
\end{figure}

The electronic density of states (DOS) of two K (2$\times$2)
overlayers on HOPG as well as the DOS of the separated subsystems
are shown in Fig. \ref{DOS}. The calculations were performed with a
13$\times$13$\times$1 Monkhorst-Pack {\bf k}-point mesh
corresponding to 87 explicit {\bf k}-points in the lateral
dimension. The graphite DOS (lower panel) shows typical features: a
steep rise at $-20$~eV due to the 2D character of graphite, a dip at
$-13$~eV after the first two $\sigma$ bands, a large peak at $-6.5$~eV
followed by a shoulder, and a zero weight
and zero gap at the Fermi energy.\cite{Cha91,Cha92,Ryt04} Despite
the charge transfer and redistribution, the characteristic features
of the graphite DOS remain as the K film is added (upper panel). The
sharp peak at $-16$~eV corresponds to the K $3p$ semicore electrons,
and important changes are seen near the Fermi energy, where the system
shows a deviation from the "V shaped" profile of HOPG. The DOS of
the separated K film (lower panel, dashed line) shows
terrace-like features in agreement with a free-electron model for a
2D metal, and it has a finite weight at the Fermi energy (second
terrace). The DOS of the whole system can then be
understood as a sum of its constituents, where K forms a decoupled
metallic layer that has signs of quantization in the perpendicular
direction.

Recent photoemission experiments and DFT calculations of Pivetta
{\it et al.} reported that the adsorption of alkali atoms induces a
gap opening in the surface electronic structure of HOPG,
\cite{Piv05} and the calculated DOS for a Na adatom [(5$\times$5)
construction, three GR layers] showed a gap of 0.15~eV in DOS. This 
is 0.4~eV below the Fermi level and in very good agreement with
experiment. It was suggested that the origin of this feature
lies in the charge transfer to the topmost GR layers (each is 
a 2D semimetal and accepts varying amounts of charge), and this
causes perturbations in the graphite electronic band structure.
We have analyzed the electronic band structure of Na
adatom on HOPG (not shown) and find a similar gap (0.17~eV) at
$-0.55$~eV. The slight difference between the numerical values is 
probably due to the different coverage [(4$\times$4) construction
in the present work].

\section{Conclusion}

We have made a systematic study of alkali metal atoms, dimers and
monolayers on HOPG using a DFT method employing a periodic slab
geometry ({\bf{k}}-points) and an extensive plane wave basis set.
This method enables us to capture the band structure of graphite and 
model a real substrate, unlike other that use a "cluster model". Our 
previous experience with Na clusters on HOPG\cite{Ryt04} has shown that 
a large basis set and a substrate of three GR layers are needed in 
order to describe the details of adsorbate-substrate interaction
accurately. The calculations are demanding in terms of both CPU time 
and memory.

In order to simulate separated adatoms, a (4$\times$4) coverage 
has been chosen (adatom separation 9.84 {\AA}), resulting in a model
substrate of 32 C atoms per GR layer (altogether 96 C atoms). This 
coverage is still far from the "real" dispersed phase with alkali-alkali 
distances of several nanometers, where the interaction is dominated by 
the Coulomb interaction with the positively charged adatoms,
and we expect a slight shift in the adatom adsorption energies as the 
coverage is decreased. The same model substrate has been applied for 
the other adsorbates as well, in order to permit a detailed comparison 
with the calculated numerical values. The bond lengths of Rb and Cs 
dimers were too large for the lateral dimension of the simulation box, 
and they were not considered in this study.

In general, alkali metal adatoms bind at the hollow site of the hexagonal 
graphite surface with adsorption energies ranging between $0.55-1.21$~eV. 
The ordering of binding energies is Li$>$Cs$\ge$Rb$\ge$K$>$Na, and the 
weak binding of Na compared with Li and K has been reported by earlier
studies.\cite{Zhu04,Val06} The mapping of the adatom locations on HOPG 
shows that Li has a moderate diffusion barrier of 0.21~eV, whereas the 
larger alkalis are relatively mobile with diffusion barriers of 
$0.02-0.06$~eV. The results for the (2$\times$2) monolayers show that the 
"dispersed" phase is more stable except for Na. This result is 
particularly important for the larger alkalis (K, Rb, Cs), which form
(2$\times$2) MLs as the coverage increases above a certain critical
value.\cite{Car05} The ordering K$>$Rb$>$Cs does not conform to 
the bulk nearest-neighbor distances, where Rb has the closest value of 
4.84~\AA\ compared with the monolayer value (4.92~\AA, note the change 
from 3D to 2D). The low formation energy of Cs may be due to the 
compressed Cs-Cs bonds, since the bulk nearest-neighbor distance is 
5.24~\AA. Despite the pronounced decoupling from the surface, Na monolayer
has a comparable formation energy to the adatom case, which is consistent 
with the clustering processes found experimentally and theoretically. 
\cite{Bre01,Bre03,Ryt04}

The amount of charge transfer and the nature of the alkali-HOPG interaction 
have been subjects of debate for many years. Our systematic study of the 
electron density redistribution upon adsorption ($\Delta\rho_\bot$) 
suggests that the charge loss is $0.4-0.5 e$ per alkali adatom. 
Furthermore, values of the order of $0.1 e$ per atom have been observed for 
(2$\times$2) MLs, indicating a decoupling effect between the alkali layer 
and HOPG. The analysis of the charge density redistribution in the 
substrate shows that the accepted charge resides mainly in the topmost 
graphene layer. We note that the calculated $\Delta q$ values depend on the 
method of evaluation, as recently shown by Valencia {\it et al.} for
several approaches ($\Delta\rho_\bot$, L\"owdin charge analysis, Bader 
atoms, Voronoi deformation charge).\cite{Val06} We prefer the simplicity of 
a method based on $\Delta\rho_\bot$, which can be used for different alkali 
metals without adjustable parameters.

The visualizations of the Li and K overlayers shed light to the
alkali-HOPG interaction. In the case of two K (2$\times$2) MLs, K 
forms a metallic film that donates charge to the $\pi$-bands of the 
substrate. The corresponding region of charge depletion is restricted to 
below the metal layer, and ELF further validates the picture of a decoupled 
2D metal (ionic bonding). The situation is slightly different for Li 
(2$\times$2) ML, as there is a pronounced charge accumulation towards the 
six nearby C atoms, and the Li atoms appear more localized. This is to be 
expected, because the Li-Li separation is far from that in bulk (3.04~\AA). 
ELF indicates that there is no chemical bonding between the adsorbate and
HOPG, but a a trace of polarization can be observed in the nearby C
atoms.

Finally, we turn to an obvious question raised by this article: Why does 
Na bind the weakest among all the alkali metals? The quantum chemical 
calculations of cation/$\rm\pi$ complexes by Tsuzuki {\it et al.}\cite{Tsu01} 
showed that polarization (induction) and electrostatic interactions are the 
major sources of attraction, and the polarization dominates binding with 
Li$^+$ and Na$^+$. Furthermore, the polarization energy was estimated to be
proportional to $R^{-4}$ for cation-benzene complexes ($R$ is the separation 
from the center of the benzene ring).\cite{Tsu01} If we study the 
corresponding $d_\bot$ values in Table \ref{tab1}, we see that the alkali 
separation is 0.6~\AA\ larger for Na than for Li. On the other hand, the 
slight decrease in IP (5.39 $\to$ 5.14 eV) indicates that the cost of 
transferring charge from the alkali towards HOPG (work function 
$\sim 4.6$~ eV) is not lowered significantly. A similar comparison between Na 
and K shows a significant drop in IP (0.8 eV), but a small increase in 
$d_\bot$. We expect that K binds more strongly than Na, and the high 
ionization potential and the relatively large surface separation (atomic 
radius) of Na are then responsible for the weak binding on graphite.

\section{Acknowledgments}

Financial support from the Academy of Finland and from the European
Community project ULTRA-1D (NMP4-CT-2003-505457) is acknowledged.
The calculations were performed on IBM-SP4(+) computers at the
Center for Scientific Computing (CSC), Espoo, Finland, and at the
John von Neumann Institute for Computing (NIC), Forschungszentrum
J\"ulich, Germany. We thank R.O. Jones for a critical reading of the 
manuscript.


\eject


\begin{table}
\caption{
Alkali metal atoms, dimers, and (2$\times$2) monolayers on graphite (0001). The formation
energy per atom ($\Delta E$) consists of the adsorption energy ($\Delta E_\bot$) and
adsorbate binding energy ($E_b$). The diffusion barrier ($E_{diff}$) corresponds to
a C-C bridge location as the adatoms prefer the hollow site. The vertical displacement
from the uppermost graphene plane ($d_\bot$), the bond distance of alkali dimers
($R_{dim}$), and the estimated charge transfer ($\Delta q$) are reported also.
$\Delta q<0$ implies a charge transfer to the substrate (in
electrons/9.84$\times$8.53 {\AA}$^2$).}
\label{tab1}
\begin{center}
\vspace{2pt}
\begin{tabular}{c c c c c c c c}
\hline
Adsorbate & $\Delta E$ (eV) & $\Delta E_\bot$ (eV) & $E_b$ (eV) &
 $E_{diff}$ (eV) & $d_\bot$ (\AA) & $R_{dim}$ (\AA) & ${\Delta}q$ \\
\hline
Li atom & 1.21 & 1.21 & - & 0.21 & 1.84 & - & -0.42 \\
Li$_2$ $^\dagger$ & 0.89/0.81 & 0.40/0.33 & 0.50 & - & 2.03/2.07 & 2.95/2.90 & - \\
        & & & & & & (2.68) & \\
(2$\times$2) ML & 0.79 & 0.39 & 0.40 & - & 2.02 & - & -0.42 \\
\hline
Na atom & 0.55 & 0.55 & - & 0.06 $\ddag$ & 2.45 & - & -0.44 \\
Na$_2$ $^{\dagger , \ddagger}$ & 0.49/0.55 & 0.14/0.20 & 0.35 & - & 3.95/3.05 & 3.07/3.16 & - \\
        & & & & & & (3.05) & \\
(2$\times$2) ML & 0.55 & 0.16 & 0.40 & - & 3.16 & - & -0.34 \\
\hline
K atom & 0.99 & 0.99 & - & 0.05 & 2.72 & - & -0.53 \\
K$_2$ $^\dagger$ & 0.74/0.62 & 0.40/0.28 & 0.34 & - & 2.91/2.93 & 4.65/4.49 & - \\
       & & & & & & (4.04) & \\
(2$\times$2) ML & 0.81 & 0.25 & 0.56 & - & 3.17 & - & -0.50 \\
2ML    &0.82 & 0.13  & 0.68 & - & 3.11/7.01 & - & -0.48 \\
\hline
Rb atom & 1.02 & 1.02 & - & 0.03& 3.15 & - & -0.48 \\
(2$\times$2) ML & 0.73 & 0.28 & 0.46 & - & 3.53 & - & -0.48 \\
\hline
Cs atom & 1.04 & 1.04 & - & 0.02 & 3.44 & - & -0.50 \\
(2$\times$2) ML & 0.57 & 0.35 & 0.22 & - & 3.75 & - & -0.52 \\
\hline
\end{tabular}
$^\dagger$The first value is for the horizontal and the second for the vertical
orientation of a dimer. The number in parentheses is the gas phase value. \\
$^\ddagger$Calculated in an orthorhombic box. The geometry optimization is
done with the 2$\times$2$\times$1 {\bf{k}}-point mesh, and the energy is
calculated with the 5$\times$5$\times$1 {\bf{k}}-point mesh.
\end{center}
\end{table}


\begin{table}
\newpage
\caption{Comparison between different DFT methods.}
\label{tab2}
\begin{center}
\vspace{2pt}
\begin{tabular}{l l l l l}  \hline

Adsorbate    & Functional & $\Delta E$ (eV) & $d_\bot$ (\AA) & Reference                           \\
\hline
Li atom         & PBE         & 1.21        & 1.84        & This work                              \\
                & PBE (LDA)   & 1.10 (1.68) & 1.71 (1.63) & Ref. \onlinecite{Val06}, slab model    \\
                & B3LYP       & 1.36        & 1.71        & Ref. \onlinecite{Zhu04}, cluster model \\
                & LDA         & 1.60        & 1.64        & Ref. \onlinecite{Kha04}, slab model    \\
(2$\times$2) ML & PBE         & 0.79        & 2.02        & This work                              \\
                & LDA         & 0.93        & 1.64        & Ref. \onlinecite{Kha04}, slab model    \\
\hline
Na atom         & PBE         & 0.55        & 2.45        & This work                              \\
                & PBE         & 0.50 (0.69) & 2.34 (2.42) & Ref. \onlinecite{Val06}, slab model    \\
            & B3LYP       & 0.72        & 2.10        & Ref. \onlinecite{Zhu04}, cluster model \\
                & PW91        &             & 2.32        & Ref. \onlinecite{Piv05}, slab model    \\
\hline
K atom          & PBE         & 0.99        & 2.72        & This work                              \\
                & PBE (LDA)   & 0.88 (1.12) & 2.65 (2.70) & Ref. \onlinecite{Val06}, slab model    \\
                & BP86 (LDA)  & 1.49 (1.67) & 2.81 (2.73) & Ref. \onlinecite{Lou00}, cluster model \\
            & B3LYP       & 1.06        & 2.51        & Ref. \onlinecite{Zhu04}, cluster model \\
                & LDA         & 0.51        & 2.79        & Ref. \onlinecite{Lam98}, slab model    \\
            & LDA         & 0.78        & 2.77        & Ref. \onlinecite{Anc93}, slab model (1 GR) \\
(2$\times$2) ML & PBE         & 0.81        & 3.17        & This work                              \\
                & LDA         & 0.98        & 2.82        & Ref. \onlinecite{Lam98}, slab model    \\
            & LDA         & 0.48        & 2.82        & Ref. \onlinecite{Anc93}, slab model (1 GR) \\
\hline
\end{tabular}
\end{center}
\end{table}

\begin{table}
\newpage
\caption{Charge transfer within the alkali-HOPG systems (in electrons/9.86$\times$8.53 \AA$^2$).
$N$ is the number alkali atoms in the simulation box and $R_{cut}$ is the adsorbate-substrate
cutoff distance. The values in parentheses are the charge transfer per alkali metal adatom.}
\label{tab3}
\begin{center}
\vspace{2pt}
\begin{tabular}{c c c c c c c}  \hline

Adsorbate       &  GR1  &  GR2  &  GR3  &  Metal              &  $N$ & $R_{cut}$ (\AA) \\
\hline
Li atom         & 0.06  & 0.09  & 0.28  & -0.42               &  1   & 1.67 \\
(2$\times$2) ML & 0.06  & 0.13  & 0.22  & -0.42 (-0.11)       &  4   & 1.48 \\
\hline
Na atom         & 0.05  & 0.09  & 0.31  & -0.44               &  1   & 1.67 \\
(2$\times$2) ML & 0.04  & 0.08  & 0.23  & -0.34 (-0.09)       &  4   & 1.61 \\
\hline
K atom          & 0.07  & 0.09  & 0.37  & -0.53               &  1   & 1.72 \\
(2$\times$2) ML & 0.06  & 0.10  & 0.29  & -0.50 (-0.13)       &  4   & 1.58 \\
2ML             & 0.09  & 0.11  & 0.28  & -0.48 (-0.12/-0.06)$^\dagger$ &  4/8$^\dagger$ & 1.58 \\
\hline
Rb atom         & 0.07  & 0.08  & 0.33  & -0.48               &  1   & 1.74 \\
(2$\times$2) ML & 0.06  & 0.09  & 0.34  & -0.48 (-0.12)       &  4   & 1.62 \\
\hline
Cs atom         & 0.07  & 0.09  & 0.35  & -0.50               &  1   & 1.79 \\
(2$\times$2) ML & 0.06  & 0.09  & 0.36  & -0.52 (-0.13)       &  4   & 1.63 \\
\hline
\end{tabular}
\end{center}
$^\dagger$ There are eight K atoms in two layers, where the lower layer (four atoms) is
in a direct contact with the substrate.
\end{table}


\begin{thebibliography}{30}
\bibitem{PES}   The energy difference between the potential energy maximum (above C atoms) 
and the saddle point (C-C bridge) is 0.03 eV for Li, and less than 0.01 eV for the other 
alkali metals.
\end{thebibliography}
\end{document}